\documentclass[aps,prl,a4paper,10pt,twocolumn,showpacs,floatfix,longbibliography,superscriptaddress,amsmath,amsfonts,amssymb,preprintnumbers,nofootinbib]{revtex4-1} 
\usepackage{graphicx}
\usepackage{dcolumn}
\usepackage{bm}
\usepackage{multirow}
\usepackage{array}
\usepackage[usenames,dvipsnames]{xcolor}
\definecolor{nblue}{rgb}{0.0, 0.0, 1.0}
\definecolor{orange}{rgb}{1.0, 0.22, 0.0}
\definecolor{green}{rgb}{0.0, 0.5, 0.0}
\usepackage[colorlinks,linkcolor=purple,urlcolor=magenta,citecolor=blue,plainpages=false,
pdfpagelabels,breaklinks]{hyperref}
\usepackage{booktabs}
\usepackage{ctable}
\usepackage{upgreek}
\usepackage{mathrsfs}
\usepackage{amssymb}
\usepackage{amsbsy}
\usepackage{color}
\usepackage{cancel}
\usepackage[bbgreekl]{mathbbol}
\usepackage{marginnote}
\usepackage{amsfonts}
\usepackage[caption=false]{subfig}

\newcommand{\bcen}{\begin{center}}
\newcommand{\ecen}{\end{center}}
\newcommand{\btab}{\begin{tabular}}
\newcommand{\etab}{\end{tabular}}
\newcommand{\bdes}{\begin{description}}
\newcommand{\edes}{\end{description}}

\newcommand{\beq}{\begin{equation}}
\newcommand{\eeq}{\end{equation}}
\newcommand{\bea}{\begin{eqnarray}}
\newcommand{\eea}{\end{eqnarray}}
\newcommand{\non}{\nonumber}

\newcommand{\bary}{\begin{array}}
\newcommand{\eary}{\end{array}}
\newcommand{\benum}{\begin{enumerate}}
\newcommand{\eenum}{\end{enumerate}}
\newcommand{\bitem}{\begin{itemize}}
\newcommand{\eitem}{\end{itemize}}

\newcommand{\bd} { \mbox{\boldmath $d$}}

\newcommand{\bk} { \bm{k} }

\newcommand{\bp} { \bm{p} }
\newcommand{\bq} { \bm{q} }
\newcommand{\br} { \boldsymbol{r}}

\newcommand{\ket}[1]{| #1 \rangle}

\newcommand{\eqn}[1] {Eqn.~(\ref{#1})}

\newcommand{\fig}[1]{Fig.~\ref{#1}}

\makeatletter

\newcommand{\Rmnum}[1]{\expandafter\@slowromancap\romannumeral #1@}
\makeatother

\newlength{\myfigwidth}
\newlength{\myhalffigwidth}
\newcommand{\calH} {\mathcal{H}}

\newcommand{\calF} {\mathcal{F}}
\newcommand{\calS} {\mathcal{S}}
\newcommand{\calZ} {\mathcal{Z}}

\newcommand{\matrixelem}[3]{\left< #1 \vphantom{#2} \left| #2
 \right| #3 \vphantom{#1} \right>}

\makeatletter
\newcommand{\thickhline}{%
    \noalign {\ifnum 0=`}\fi \hrule height 2pt
    \futurelet \reserved@a \@xhline
}
\newcolumntype{"}{@{\hskip\tabcolsep\vrule width 2pt\hskip\tabcolsep}}

\newcommand{\mylabel}[1]{\label{#1}} 

\newsavebox{\measurebox}

\begin{document}



\title{Quantitative Theory of Triplet Pairing\\in the Unconventional Superconductor LaNiGa$_2$}
\author{Sudeep Kumar Ghosh}
\email{S.Ghosh@kent.ac.uk}
\affiliation{Physics of Quantum Materials, School of Physical Sciences, University of Kent, Canterbury CT2 7NH, United Kingdom}
\author{G{\'a}bor Csire}
\email{gabor.csire@icn2.cat}
\affiliation{H. H. Wills Physics Laboratory, University of Bristol, Tyndall Avenue, Bristol BS8 1TL, United Kingdom}
\affiliation{Catalan Institute of Nanoscience and Nanotechnology (ICN2), CSIC, BIST, Campus UAB, Bellaterra, Barcelona, 08193, Spain}
\author{Philip Whittlesea}
\affiliation{Physics of Quantum Materials, School of Physical Sciences, University of Kent, Canterbury CT2 7NH, United Kingdom}

\author{James F. Annett}
\affiliation{H. H. Wills Physics Laboratory, University of Bristol, Tyndall Avenue, Bristol BS8 1TL, United Kingdom}
\author{Martin Gradhand}
\affiliation{H. H. Wills Physics Laboratory, University of Bristol, Tyndall Avenue, Bristol BS8 1TL, United Kingdom}
\author{Bal{\'a}zs {\'U}jfalussy}
\affiliation{Institute for Solid State Physics and Optics, Wigner Research Centre for Physics, Hungarian Academy of Sciences, PO Box 49, H-1525 Budapest, Hungary}
\author{Jorge Quintanilla}
\email{J.Quintanilla@kent.ac.uk}
\affiliation{Physics of Quantum Materials, School of Physical Sciences, University of Kent, Canterbury CT2 7NH, United Kingdom}

%
%


\date{\today}

\begin{abstract}

The exceptionally low-symmetry crystal structures of the time-reversal symmetry breaking superconductors LaNiC$_2$ and LaNiGa$_2$ lead to an internally-antisymmetric non-unitary triplet (INT) state as the only possibility compatible with experiments. We argue that this state has a distinct signature: a double-peak structure in the Density of States (DOS) which resolves in the spin channel in a particular way. We construct a detailed model of LaNiGa$_2$ capturing its electronic band structure and magnetic properties {\it ab initio}. The pairing mechanism is described {\it via} a single adjustable parameter. The latter is fixed by the critical temperature $T_c$ allowing parameter-free predictions. We compute the electronic specific heat and find excellent agreement with experiment. The size of the ordered moment in the superconducting state is compatible with zero-field muon spin relaxation experiments and the predicted spin-resolved DOS suggests the spin-splitting is within the reach of present experimental technology. 

\end{abstract}



\maketitle

The superconducting state is a condensate of electron pairs characterized by an order parameter $\Delta$. Usually $\Delta$ is a complex scalar, its phase being a manifestation of spontaneously-broken gauge symmetry. This is responsible for the macroscopic quantum coherence underpinning quantum devices such as superconducting qubits~\cite{Devoret2004} and SQUIDs~\cite{Fagaly2006}. On the other hand, in so-called ``unconventional'' superconductors additional symmetries may be broken leading to more complex order parameters with extra degrees of freedom. Of all the features of unconventional superconductors, broken time-reversal symmetry (TRS) is perhaps the most surprising one as it challenges our view of superconductivity and magnetism as antagonistic states of matter. In spite of this, the phenomenon has been detected in numerous systems using zero-field muon spin rotation/relaxation ($\mu$SR). Prominent examples include (U, Th)Be$_{13}$~\cite{Heffner1990}, Sr$_2$RuO$_4$~\cite{Luke1998}, UPt$_3$~\cite{Luke1993}, (Pr, La)(Ru, Os)$_4$Sb$_{12}$~\cite{Aoki2003, Shu2011}, PrPt$_4$Ge$_{12}$~\cite{Maisuradze2010}, LaNiC$_2$~\cite{Hillier2009}, LaNiGa$_2$~\cite{Hillier2012,Weng2016}, SrPtAs~\cite{Biswas2013}, Re$_6$(Zr, Hf, Ti)~\cite{Singh2014,Singh2017,Shang2018,Singh2018}, Lu$_5$Rh$_6$Sn$_{18}$~\cite{Bhattacharyya2015} and  La$_7$(Ir, Rh)$_3$~\cite{Barker2015,Singh2018a}. Many of these systems have other unconventional features, while in some cases an independent, direct observation of broken TRS has been made:  optical Kerr effect measurement in Sr$_2$RuO$_4$~\cite{Xia2006} and UPt$_3$~\cite{Schemm2014}, and bulk SQUID magnetization measurement in LaNiC$_2$~\cite{Sumiyama2015}.

 Unfortunately it has been difficult to establish the structures of order parameters of these superconductors. This is because, on the one hand, our knowledge of the electron pairing mechanism is not sufficiently detailed to make a prediction. On the other hand, their crystal structures tend to be highly symmetric, leading to many different possible ways of breaking TRS, which  limits our ability to work by elimination. TRS-breaking superconductivity requires a degenerate instability channel~\cite{Annett1990,Sigrist1991} which, for a uniform superconductor, must come from a multi-dimensional irreducible representation (irrep) of the point group of the crystal. As an example, the point group of Sr$_2$RuO$_4$ is $D_{4h}$, which leads to 22 possible order parameters breaking TRS~\cite{Annett1990,Sigrist1991}: 20 under the assumption of weak spin-orbit coupling (SOC) and two more in the strong-SOC limit. The family formed by LaNiC$_2$~\cite{Hillier2009} and LaNiGa$_2$~\cite{Hillier2012} are an exception to this rule, as their crystal structures have exceptionally low symmetry. Their crystal point groups only have four irreps, all of them one-dimensional. This precludes TRS breaking in the strong-SOC case and leaves only four possible pairing states, all of them non-unitary triplets~\cite{Quintanilla2010,Hillier2012}. One additional complication is the multi-band nature of these systems: two~\cite{Subedi2009} and five~\cite{Singh2012} bands cross the Fermi level of LaNiC$_2$ and LaNiGa$_2$, respectively. In fact, both systems show thermodynamic properties that can be fitted with a model assuming fully-gapped, 2-band superconductivity~\cite{Chen2013,Chen2013A,Weng2016}. This is inconsistent with the line nodes implied by the earlier symmetry analyses~\cite{Quintanilla2010,Hillier2012}. On the other hand the 2-band model does not predict TRS breaking. To resolve the discrepancy it was proposed that only an internally-antisymmetric non-unitary triplet pairing (INT) state is compatible with the experimental observations~\cite{Quintanilla2010,Hillier2012,Weng2016,nonunitary}. Here we show that such a state has a very distinct experimental signature: a double-peak structure in the Density of States (DOS) which resolves in the spin channel. We construct a model of LaNiGa$_2$ capturing detailed electronic band structure {\it ab initio}, with the pairing interaction in the INT state reduced to a single, adjustable parameter. The known value of the critical temperature $T_c$ fixes this single parameter, allowing us to make parameter-free predictions. We obtain the electronic specific heat and find an excellent agreement with experiment~\cite{Weng2016}. We compute the spin-resolved DOS having a double-peak structure with each peak corresponding to a single spin channel. We find that the splitting $\sim 0.2 {\rm meV}$ ---within the reach of present experimental technology.

The triplet pairing in the INT state relies heavily on the inter-band pairing, which enables an isotropic gap function and equal-spin pairing breaking TRS~\cite{Weng2016}.  The Cooper pair wave function is symmetric in the crystal momentum and spin channels but it is anti-symmetric with respect to the orbital degree of freedom. Recent studies~\citep{Dai2008,Weng2016,Nomoto2016,Brydon2016,Yanase2016,Nica2017,Agterberg2017,Brydon2018,Huang2019,
Ramires2019,Hu2019,Lado2019,li2012} in several materials, including the Iron based superconductors, half-Heusler compounds, UPt$_3$ and Sr$_2$RuO$_4$, have also pointed out the importance of internal degrees of freedom of electrons (coming from, for example, sublattice or multiple orbitals) in determining the pairing symmetries of superconducting ground states.      

A convenient toy model of low-energy excitations in the INT state proposed in Ref.~\cite{Weng2016} is provided by the following Bogoliubov-de Gennes (BdG) Hamiltonian: 
\beq
\mathcal{H}=
\begin{pmatrix}
  \mathcal{H}_0(\bk)  & \hat{\Delta}                \\ 
  \hat{\Delta}^{\dagger}       & - \mathcal{H}_0(\bk) 
\end{pmatrix}.
\label{HBdG}
\eeq
Here $\mathbf{k}$ is the crystal momentum of the excitation, 
\beq\mylabel{eqn:H0mat}
\mathcal{H}_0(\bk) = \mathbb{1}_2 \otimes \begin{pmatrix}
\epsilon_0(\bk) -\mu - s & \delta 
\\ \delta 
& \epsilon_0(\bk) -\mu + s
\end{pmatrix}
\eeq
is the normal-state, single-electron Hamiltonian with the chemical potential $\mu$ and 
\beq
\hat{\Delta} = i(\bd.\pmb{\sigma})\sigma_y \otimes i \tau_y\eeq 
represents the pairing potential. In the tensor products, the first sector represents the spin channels $\sigma=\uparrow,\downarrow$ while the second represents the two orbital channels. For the purpose of initial discussion, we have assumed a very simple band structure with two bands labeled by $m = +$ and $-$, one emerging from each orbital, that are related by a rigid energy shift $2s$ and with a $\mathbf{k}-$independent hybridization factor $\delta$. The pairing matrix describes $\mathbf{k}-$independent triplet pairing but is antisymmetric in the orbital channel in order to ensure the fermionic antisymmetry of the Cooper pair wave function~\cite{Dai2008,Weng2016}. Here, $\pmb{\sigma}$ and $\pmb{\tau}$ are the vectors of Pauli matrices in the spin and orbital sectors respectively. Writing the triplet $d$-vector in the form $\bd = \Delta_0 \pmb{\eta}$, where $|\pmb{\eta}|^2 = 1$ and $\Delta_0$ is a pairing amplitude, the nonunitarity of the triplet state is characterized by a nonzero real vector $\bq = i (\pmb{\eta} \times \pmb{\eta}^*)$ which in general has $|\bq| \leq |\pmb{\eta}|^2 = 1$. 

Diagonalizing $\mathcal{H}$ yields the quasi-particle spectrum  
$E_{\bk}$
shown, for a particular choice of parameters, in Fig.~\ref{fig:spectrum}(a). The plot is representative of cases where $s,\delta \ll \Delta_0$. This is the physically-relevant regime {for the toy model} as in a mean-field picture the pairing amplitude ($\Delta_0$) has to be able to overcome the band splitting $\sim \delta,s$. {This unrealistic requirement is relaxed when the band splitting is allowed to be k-dependent, as in the more detailed model discussed below.} As indicated in the plot each excitation has well-defined band and spin indices. The Bogoliubov bands are paired up, with each member of the pair sharing the spin index but differing in the band index. The corresponding DOS is displayed in Fig.~\ref{fig:spectrum}~(b) and (c). Here we have introduced two different levels of broadening to simulate different experimental resolutions in the two figures. The DOS is fully-gapped, with four pairs of coherence peaks that are grouped in two doublets, depending on the level of broadening. Crucially, the \emph{spin-resolved} DOS shows only one of the two doublets in each spin channel. This qualitative feature distinguishes this double-peak structure from that which would be obtained, for example, in a multi-band superconductor. The observation of such a spin-resolved feature would provide definitive proof of the INT state.

\begin{figure}
{
\includegraphics[width=0.48\textwidth]{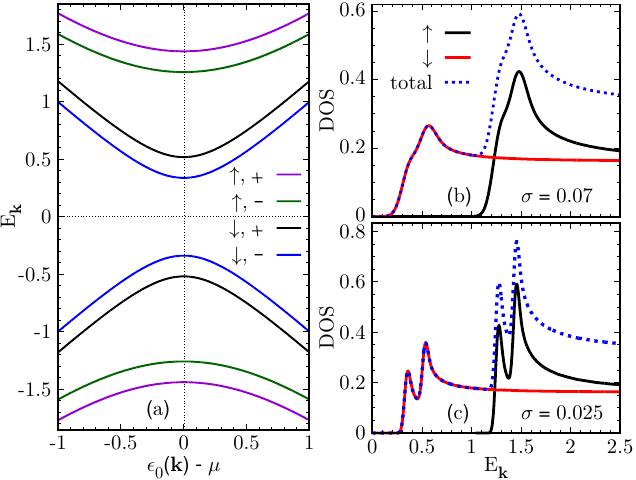}
}
\caption{Properties of quasiparticles in the INT state with $s=0.05$, $\delta = 0.075$, $|\Delta_0| = 1$ and $|\bq|= \sqrt{2/3}$ in arbitrary units. (a) Quasiparticle spectrum for the $+$ and $-$ bands for $\uparrow$ and $\downarrow$ spins. (b) and (c) show the corresponding DOS calculated from this spectrum using the same parameters. The DOS features have been artificially broadened by convolution with a Gaussian of width $\sigma=0.07$ in (b) and $\sigma=0.025$ in (c).}
\mylabel{fig:spectrum}
\end{figure}

An analytical formula for $E_{\bk}$ can be easily obtained in the limits $s \to 0$ or $\delta \to 0$. In either case, the result is  
\beq
E_{\bk} = \pm \left[\pm a + \sqrt{\{\epsilon_0(\bk)-\mu\}^2 + |\Delta_0|^2 (1 \pm |\bq|)} \right],
\eeq 
where $a=\delta$ or $s$, respectively. {Note that both $\delta$ and $s$ play similar roles. This shows that our toy model does not rely on the two bands being orthogonal.} {The above formula can be used} to estimate the ratio between the gaps in the energy spectrum for spin-up ($\mathcal{E}_{\uparrow\uparrow}$) and spin-down ($\mathcal{E}_{\downarrow\downarrow}$) quasiparticles. In the limit ${a}\ll\Delta_0$, it is
\beq
\frac{\mathcal{E}_{\uparrow\uparrow}}{\mathcal{E}_{\downarrow\downarrow}}
=
\frac{\sqrt{1+|\bq|}}{\sqrt{1-|\bq|}}
=1-|\bq|+ O({|\bq|^2}).
\eeq

The above toy model assumes that an isotropic, equal-spin pairing potential can lower the free energy in spite of the need for it to breach the energy gap between the bands. We explicitly show this by considering the toy many-body Hamiltonian
\beq
\hat{H}=\sum_{\bk}\hat{\Psi}^{\dagger}_{\bk}\mathcal{H}_0(\bk)\hat{\Psi}_{\bk}+
\hat{H}_I.
\label{eqn:toy_H}
\eeq
Here $\Psi_{\bk}=(\hat{c}_{\bk,+,\uparrow},\hat{c}_{\bk,+,\downarrow},\hat{c}_{\bk,-,\uparrow},\hat{c}_{\bk,-,\downarrow})$, where $\hat{c}_{\bk,m,\sigma}$ creates an electron in the $m^{\rm th}$ band with crystal momentum $\bk$ and spin $\sigma$. The single-electron Hamiltonian $\mathcal{H}_0(\bk)$ is given in Eq.~(\ref{eqn:H0mat}) where for simplicity we take $\delta=0$ and $\epsilon_0(\bk) = -2t[\cos(k_x) + \cos(k_y)]$. We consider an on-site, inter-orbital, equal-spin pairing interaction proposed in Ref.~\cite{Weng2016}, which can be written~\cite{origin} as
\beq
\hat{H}_I=-U \sum_{\bk,\bk',\sigma} c^{\dagger}_{\bk,+,\sigma}c^{\dagger}_{-\bk,-,\sigma}c_{-\bk',-,\sigma}c_{\bk',+,\sigma}
\label{eqn:HI}
\eeq
with $U>0$ being the effective attraction strength. A standard mean-field treatment of this model (see Supplemental Material) yields the phase diagram shown in Fig.~\ref{fig:ESP_properties}. In the limit $s\to 0$, the theory is formally equivalent to two copies of a BCS theory, but with the band index $m$ playing the role of the spin index (one copy corresponding to each value of the real spin). For finite $s$, a critical interaction strength $U_c$ is necessary for the critical temperature $T_c$ to be finite, but for $U\gg U_c$ the results are very similar to the case $s\to 0$. This is confirmed by the inset, showing the temperature-dependence of the pairing amplitude $\Delta_0$. While the superconducting transition can be of first order and even re-entrant (not shown; see Supplemental Material) for a very narrow window $U \gtrsim U_c$, and displays some shoulders for slightly larger $U$, BCS-like behavior is recovered for $U \gg U_c$.  

\begin{figure}[!ht]
{
\includegraphics[width=0.48\textwidth]{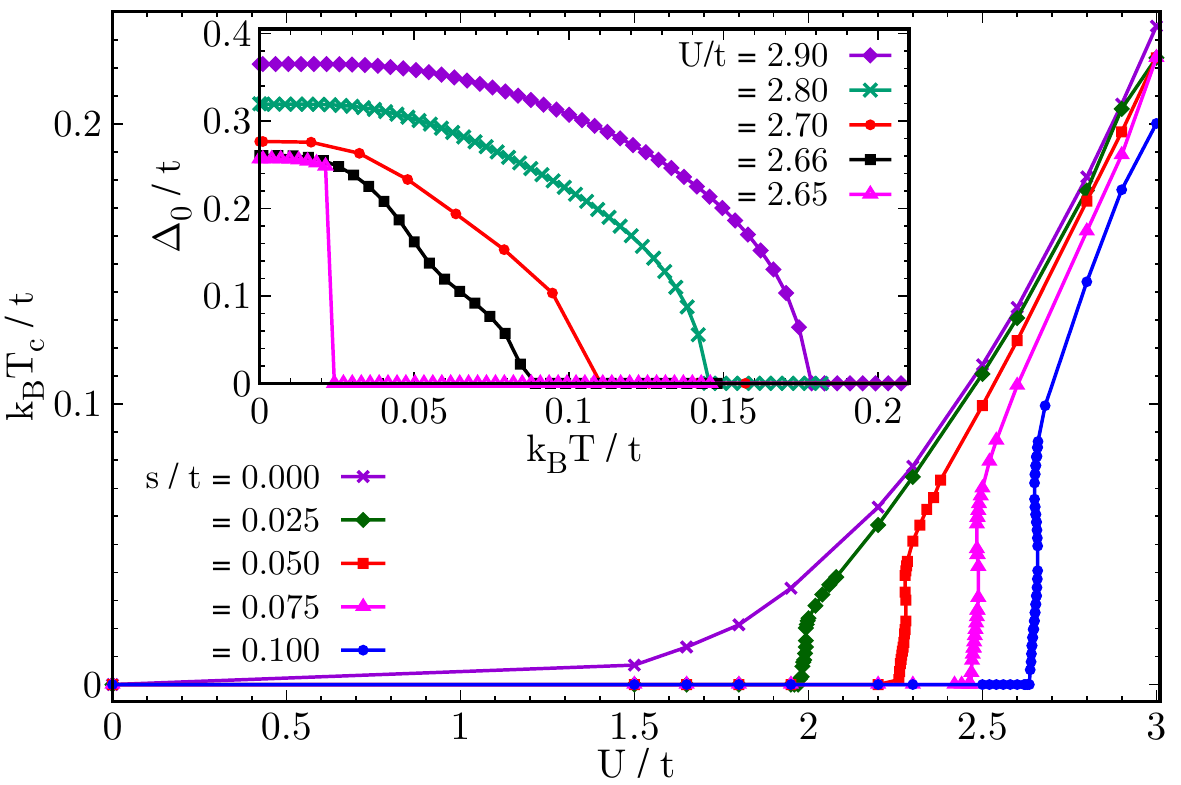}
}
\caption{Superconducting phase diagram of the toy model in Eq.~(\ref{eqn:toy_H}). $T_c$ is the critical temperature where inter-orbital, equal-spin pairing sets in. Each curve shows the dependence of $T_c$ on the interaction strength $U$ for a different value of the band splitting $2s$, as indicated. The inset shows the temperature-dependence of the pairing amplitude $\Delta_0$ for the largest splitting $s/t=0.1$ and a few values of $U$ just above the critical value $U_c$ at which $T_c$ becomes finite. 
}
\mylabel{fig:ESP_properties}
\end{figure}

The above simple calculation shows that an equal-spin pairing potential can, in principle, breach a band gap to lead to a fully-gapped triplet pairing state. On the other hand our simple mean field theory yields $\pmb{\eta}=(0,1,0)$, i.e. a unitary triplet pairing state with $\bq = 0$. A more realistic theory must treat the pairing and exchange fields on equal footing. Symmetry arguments~\cite{Hillier2012} show that any triplet instability leads to a subdominant magnetisation, which lowers the free energy of the non-unitary state with $|{\bf q}| \neq 0$~\cite{Hillier2012,Miyake2014}. We now therefore build a more sophisticated, realistic description which not only incorporates an accurate description of the exchange field, but also a realistic prediction of the normal-state electronic structure of LaNiGa$_2$, making quantitative predictions of superconducting properties possible.

\begin{figure*}[!t]
{
\includegraphics[width=0.998\textwidth]{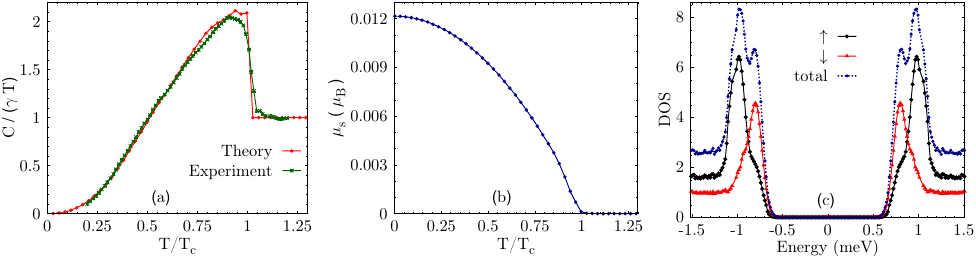}
}
\caption{(Color online) Properties of the superconducting ground state corresponding to a phenomenological inter-orbital equal-spin pairing interaction between the (d$_{z^2}$ - d$_{xy}$) orbitals in the 3d sector of the Ni atom in LaNiGa$_2$. (a) Variation of the specific heat with temperature ($\gamma$ is the Sommerfeld coefficient). We note an excellent agreement between the theoretical result and the experimental data taken from Ref.~\cite{Weng2016}. (b) Variation of the spontaneous magnetic moment ($\mu_s$) as a function of temperature. A clear increase in the magnetic moment below $T_c$ is an indication of the imbalance between two spin-species due to migration of Cooper pairs. (c) Spin-resolved density of states (arbitrary units) of the Bogoliubov quasi-particles as a function of energy. We note that the two coherence peaks correspond to up and down species of Cooper pairs leading to two gaps in the quasi-particle spectrum.}
\mylabel{fig:DFT_results}
\end{figure*}

Density-Functional Theory (DFT) in the Local-density approximation (LDA) shows that LaNiGa$_2$ is a multi-band superconductor with several bands crossing the Fermi level giving rise to multiple Fermi surface sheets~\cite{Singh2012}. None of the bands can be obtained from one another through a simple rigid shift as in our toy model. There are, however, several regions within the Brillouin zone where the pairs of Fermi surface sheets are parallel {and very close} to each other{, that is, nearly degenerate} (see Supplemental Material). Moreover, 
the five bands have mixed Ni $3$d, La $5$d and Ga $4$p characters. 
As a result, the Fermi surfaces have strong orbital degeneracy. 
To capture these details, we adopt a semi-phenomenological strategy~\cite{Gyorffy1998}.
We consider the relativistic version of the BdG Hamiltonian in Eq.~(\ref{HBdG})
together with the realistic LDA band structure and a phenomenological pairing interaction of the type given by Eq.~(\ref{eqn:HI}). This leads to the Kohn-Sham-Dirac-BdG Hamiltonian~\cite{Csire2018}
\beq
 \mathcal{H}_{DBdG} = 
  \left[ {\begin{array}{cc}
   H_D & \hat{\Delta}(\br)  \\
   \hat{\Delta}^{\dagger}(\br)  & -H^*_D \\
  \end{array} } \right]
\eeq
where $H_D$ is the effective normal state Dirac Hamiltonian given by 
\beq
H_D = c \bp \hat{\pmb{\alpha}}_1 + (\hat{\pmb{\alpha}}_2 - \mathbb{1}_4)c^2/2 + (V_{eff}(\br) - E_F)\mathbb{1}_4 + \mathbf{B}_{eff}(\br)\hat{\pmb{\alpha}}_3.
\eeq
Here, $\hat{\pmb{\alpha}}_1 = \hat{\sigma}_x \otimes \hat{\pmb{\sigma}}$, $\hat{\pmb{\alpha}}_2 = \hat{\sigma}_z \otimes \mathbb{1}_2$ and $\hat{\pmb{\alpha}}_3 = \mathbb{1}_2 \otimes \hat{\pmb{\sigma}}$ with $\hat{\pmb{\sigma}}$ being the Pauli matrices and $\mathbb{1}_n$ being the identity matrix of order $n$.
$V_{eff}(\br)$ and $\mathbf{B}_{eff}(\br)$ are the effective electrostatic potential and the effective exchange-field, respectively.
$\hat{\Delta}(\br)$ is the $4\times4$ pairing potential matrix due to the four component Dirac spinors. Requiring self-consistency in the electrostatic potential, exchange-field and pairing potential, the solution is provided by our recently developed method~\cite{Csire2018} which generalizes the Korringa-Kohn-Rostoker (KKR) formalism. Within the KKR formalism intra-orbital and inter-orbital pairings could be described both in the singlet and triplet channel by transforming the $(L,\sigma)$ representation of the pairing potential into a relativistic basis set, where $L$ refers to the real spherical harmonics assuming that the $z$ direction is perpendicular to the layered structure of LaNiGa$_2$. The technical details are given in the Supplemental Material.

It is important to note that the ground state does not show ferromagnetism in the normal state, although there is a significant contribution from the Ni $3d$ states at the Fermi level~\cite{Singh2012}. Since it is known that Hund's rule coupling plays an important role on the Ni atoms~\cite{Okabe1997,Georges2013}, and on the other hand Hund's coupling can also produce local pairing~\cite{Han2004}, therefore, it is physically reasonable to assume an inter-orbital equal-spin pairing involving two orbitals on the same Ni atom. We describe this by a two-body onsite attractive interaction $U_{L,L'}$ between electrons with equal spins in only two of these orbitals ($L\neq L'$) with the pairing potential satisfying the self-consistency equation: $\Delta_{L\sigma,L'\sigma}(\br) = U_{L,L'} \chi_{L\sigma,L'\sigma}(\br)$ where $\chi_{L\sigma,L'\sigma}(\br)$ is the corresponding pairing amplitude. Since all of the Ni $d$ orbitals contribute to the density of states at the Fermi level, there are $10$ possible pairing models within this approach. Only one of the $10$ possible combinations, namely pairing between d$_{z^2}$ and d$_{xy}$, yields a fully-gapped quasi-particle spectrum (all the other possibilities have nodes on at least one of the Fermi surface sheets-- see Supplemental Material). The strength $U_{d_{z^2},d_{xy}}$ of the interaction between these two orbitals is the \emph{only} adjustable parameter in our theory, to be fixed by requiring $T_c$ to be the same as in experiments~\cite{Weng2016}. Then we can make parameter-free predictions of observable properties of the system. The requirement that $T_c$ is the same as in experiments leads to $U_{d_{z^2},d_{xy}} = 0.65$ eV which is comparable to the values of Hund's coupling found for Ni atom~\cite{Georges2013}. This result should motivate high-pressure measurements and Dynamical Mean Field Theory studies to further explore the role of electronic correlations involving Hund's coupling. {We stress that the present attractive interaction is described by a phenomenological parameter and therefore our calculation cannot directly address the question of its origin.} 

Having fixed our single parameter, we can now make parameter-free predictions. We first compute the specific heat of LaNiGa$_2$ as a function of temperature by evaluating the temperature dependence of the quasi-particle DOS self-consistently~\cite{Gyorffy1998,Csire2018A}. It is shown in \fig{fig:DFT_results}(a) comparing to the corresponding experimental data from Ref.~\cite{Weng2016}. The agreement is excellent, suggesting that the observed two-gap behavior of this curve is consistent with our equal-spin, inter-orbital pairing model. 

The solution of the self-consistency equations reveal a charge imbalance between $\uparrow \uparrow$ ($67\%$) and $\downarrow \downarrow$ ($33\%$) triplet components on the Ni atom. The migration of Cooper-pairs from the minority $\downarrow\downarrow$ state to the majority $\uparrow \uparrow$ state is expected to generate a finite magnetization. Since our pairing interaction $U_{d_{z^2},d_{xy}}$ is spin independent (hence preserves TRS), we have found a spontaneous TRS breaking in the INT state, which is a perfect analogue of a ferromagnetic transition in a normal-state DFT calculation. The pairing-induced spontaneous magnetization ($\mu_s$) is shown in \fig{fig:DFT_results}(b). The magnetic moment is expected to vary linearly very close to $T_c$~\cite{Hillier2012}, however this behavior is hard to resolve here due to demanding numerical accuracy near $T_c$. We can estimate the size of the internal magnetic field at zero temperature using $\mu^0_s$, the value of $\mu_s$ at $T=0$, as $B_{int} \approx \frac{\mu_0 \mu^0_s}{4\pi a b c} \approx 0.3$ Gauss which is of similar order as seen in the zero-field $\mu$SR measurements~\cite{Hillier2012}.

Finally, we compute the spin-resolved quasiparticle DOS of LaNiGa$_2$ as shown in \fig{fig:DFT_results}(c). Its similarity with Fig.~\ref{fig:spectrum}(b) is striking, confirming that our DFT-KKR calculation for this material describes the same physics. The two distinct superconducting gaps are clearly visible, and the spin-resolved curves show that they correspond to different spin species. The double-peak structure of the DOS is our main prediction. We note that the splitting between the two peaks is of the order of 0.2~meV ---within the resolution of current scanning tunneling microscopy~\cite{Choi2017}, photo-emission~\cite{Fanciulli2017}, and tunneling experiments~\cite{Hirjibehedin_2014}. The crucial feature is that, unlike the case of a multi-band superconductor~\cite{Szabo2001}, the two peaks correspond to \emph{distinct spin channels}. Verifying this experimentally would thus require spin resolution~\cite{Wiesendanger2009}.

Interestingly, in Ref.~\onlinecite{Weng2016} the specific heat measurement was fitted by a phenomenological two-band model leading
to the gap values $\Delta_1=1.08 k_B T_c$ and $\Delta_2=2.06 k_B T_c$, while we find that the difference between the gaps is only around 20\%. Clearly, the difference between the two procedures comes from the fact that our first-principles based calculation included all of the five bands crossing the level. However, our main point is that the DOS is spin-polarized around the Fermi level, and the superconducting gaps correspond to different spin channels, not different bands.

We note that it is important to consider the effect of magnetic and nonmagnetic impurities on the INT state. Although this is outside the scope of the present paper, due to the two full gaps we expect the INT state to be protected from nonmagnetic impurity scattering and a version of the Anderson's theorem~\cite{anderson1959} to hold.

\textbf{Conclusions}: We showed that an unconventional superconductor in the INT state has at least two gaps, one for each spin flavour, irrespective of the number of Fermi surfaces. Instead of the traditional route of ignoring the microscopic complexity of Fermi surfaces, we consider the fully-relativistic electronic band structure of LaNiGa$_2$. {We perform fully self-consistent} computations of its observables by taking a phenomenological pairing model on the Ni atom in the INT state. {The pairing model has a single adjustable parameter fixed by the experimental value of $T_c$ of the material}. There is an excellent agreement between the computed and measured specific heat of the system. We showed that due to migration of Cooper pairs a sub-dominant order parameter, magnetization, arises spontaneously, breaking TRS {consistent with the zero-field muSR experiment}.  {The salient feature of our calculations is a double-peak structure} in the quasi-particle DOS arising from the two spin channels. {We have predicted quantitatively the splitting between the two peaks and showed that it is well} within the reach of present experimental technology and resolution. We have thus achieved a desired milestone: a quantitative theory of exotic pairing in an unconventional superconductor, namely LaNiGa$_2${, predicting a smoking-gun signature of its unconventional pairing state}.

\section{Acknowledgments}
This research was supported by EPSRC through the project ``Unconventional Superconductors: New paradigms for new materials'' (grant references EP/P00749X/1 and EP/P007392/1) and the Hungarian National Research, Development and Innovation Office under the contracts No. K115632. G.Cs. also acknowledges support from the European Union's Horizon 2020 research and innovation programme under the Marie Sklodowska-Curie grant agreement No. 754510 and thanks Tom G. Saunderson and Hubert Ebert for fruitful discussions. BU also acknowledges NKFIH K131938 and BME Nanotechnology FIKP grants. JQ thanks Silvia Ramos for useful discussions. We also thank Kazumasa Miyake for extensive discussions.

\section{Supplementary material}
In this supplemental material, we have given details of the Bogoliubov-de Gennes (BdG) formulation of the variational mean-field theory of the many-body toy model describing pairing between same spins on a square lattice in the internally-antisymmetric non-unitary triplet (INT) state. We have also given the technical details of the Kohn-Sham-Dirac BdG calculation within the Korringa-Kohn-Rostoker (KKR) method for the material LaNiGa$_2$ including the details of its Fermi surfaces.  

\subsection{Mean-field theory of the toy model}
The model many-body toy Hamiltonian given in Eq.(6) of the main text is
\beq
\hat{H}=\underbrace{\sum_{\bk}\hat{\Psi}^{\dagger}_{\bk}\mathcal{H}_0(\bk)\hat{\Psi}_{\bk}}_{\hat{\mathcal{H}}_0}+
\hat{H}_I.
\label{eqn:toy_H1}
\eeq
The interaction term describing onsite pairing between same spins but in two different orbitals is
\beq
\hat{H}_I=-U \sum_{\bk,\bk',\sigma} c^{\dagger}_{\bk,+,\sigma}c^{\dagger}_{-\bk,-,\sigma}c_{-\bk',-,\sigma}c_{\bk',+,\sigma}.
\label{eqn:HI1}
\eeq
As discussed in the main text in general we consider the two orbitals to have a rigid energy shift (bare splitting) $2s$ and a $\bk$-independent hybridization factor $\delta$. For the case of $\delta = 0$, $\hat{\mathcal{H}}_0$ takes the simple form
\beq
\hat{\mathcal{H}}_0 = \sum_{m,\bk,\sigma} [\epsilon_0(\bk) + m s -\mu] c^{\dagger}_{\bk,m,\sigma} c_{\bk,m,\sigma}.
\eeq
We will use variational mean-field theory within the Bogoliubov-de Gennes (BdG) formalism~\cite{de2018} to compute the ground state properties of the system. We consider the following mean-field ansatz for the interaction term
\bea
\hat{V}_{MF} &=& \sum_{\bk,\sigma} (\Delta_{\sigma\sigma} c^{\dagger}_{\bk,+,\sigma}c^{\dagger}_{-\bk,-,\sigma} + \mbox{h. c.}) \non\\ &+& \sum_{m,\bk,\sigma}\phi_{m,\sigma} c^{\dagger}_{\bk,m,\sigma} c_{\bk,m,\sigma}.
\eea
Note that we have used two variational mean fields $\Delta_{\sigma\sigma}$ and $\phi_{m,\sigma}$. $\Delta_{\sigma\sigma}$ describes onsite pairing between the same spins in the two different orbitals and $\phi_{m,\sigma}$ describes any spontaneous polarization that may arise in the mean-field ground state. Thus the variational mean-field Hamiltonian for the system is given by
\beq\mylabel{eqn:HMF}
\hat{\mathcal{H}}_{MF} = \hat{\mathcal{H}}_0 + \hat{V}_{MF}.
\eeq
We minimize the free energy of the system with respect to the two variational mean-fields to find the ground state properties of the system. The free energy of the system is given by
\beq
\calF = - k_B T \log(\calZ) = \langle \hat{H} \rangle - T \calS
\eeq
where, $k_B$ is the Boltzmann constant, $T$ is the temperature, $\calZ$ is the partition function of the system, $\calS$ is the entropy and the expectation value is taken with respect to the ground state of $\hat{H}$. Within mean field theory, we now approximate this free energy as follows
\bea
\calF &\approx & \langle \hat{H} \rangle_{\mathrm{MF}} - T \calS_{\mathrm{MF}} \non\\
&=& \langle \hat{H} \rangle_{\mathrm{MF}} - \langle \hat{\calH}_{\mathrm{MF}} \rangle_{\mathrm{MF}} + \langle \hat{\calH}_{\mathrm{MF}} \rangle_{\mathrm{MF}} - T \calS_{\mathrm{MF}} \non\\
\Rightarrow \calF &\approx & \langle \hat{H} - \hat{\calH}_{\mathrm{MF}} \rangle_{\mathrm{MF}} + \calF_{\mathrm{MF}} \,.\mylabel{eqn:fenergy}
\eea
Here, $\langle\,\rangle_{\mathrm{MF}}$ implies that the expectation value is evaluated in the mean field ground state. $\calS_{\mathrm{MF}}$ is the entropy and $\calF_{\mathrm{MF}} = \langle \hat{\calH}_{\mathrm{MF}} \rangle_{\mathrm{MF}} - T \calS_{\mathrm{MF}}$ is the free energy of the mean field system. We now extremize the free energy with respect to the mean fields by requiring the stationary condition
\bea
\delta \calF &=& 0\,, \non\\ 
\Rightarrow \delta\langle \hat{H} - \hat{\calH}_{\mathrm{MF}} \rangle_{\mathrm{MF}} &=& 0 \mylabel{eqn:stationary}
\eea
since by definition $\delta\calF_{\mathrm{MF}}=0$. This leads to the self-consistency equations of the mean-fields:
\bea
\Delta_{\sigma,\sigma} &=& -\frac{U}{\Omega} \chi^*_\sigma,\\
\phi_{m,\sigma} &=& -\frac{U}{\Omega} N_{-m,\sigma}.
\eea
Here, $\Omega$ is the total number of sites in the lattice and
\bea
\chi_{\sigma} &=& \sum_{\bk} \langle c^{\dagger}_{\bk,+,\sigma}c^{\dagger}_{-\bk,-,\sigma} \rangle_{\mathrm{MF}},\\
N_{m,\sigma} &=& \sum_{\bk} \langle c^{\dagger}_{\bk,m,\sigma}c_{\bk,m,\sigma} \rangle_{\mathrm{MF}}.
\eea
The free energy in \eqn{eqn:fenergy} is then given by
\bea\mylabel{eqn:fenergy1}
\calF &=& \calF_{MF} - \frac{U}{\Omega} \sum_\sigma \left(|\chi_\sigma|^2 + N_{+,\sigma} N_{-,\sigma} \right) \non \\
&-& \sum_\sigma (\Delta_{\sigma\sigma} \chi_\sigma + \mbox{h. c.}) - \sum_{m,\sigma} \phi_{m,\sigma} N_{m,\sigma}.
\eea
We diagonalize the mean-field Hamiltonian in \eqn{eqn:HMF} using standard Bololiubov transformation~\cite{de2018} to obtain
\beq
\hat{\mathcal{H}}_{MF} = \sum_{\bk,\sigma} \sum_{n=+,-} E_{n,\bk,\sigma} \gamma^{\dagger}_{n,\bk,\sigma}\gamma_{n,\bk,\sigma}
\eeq
where $\gamma^{\dagger}_{n,\bk,\sigma}$ is the creation operator of a Bogoliubov quasiparticle in the Bogoliubov band labeled by $n=+, -$ with spin $\sigma$ and momentum $\bk$ and $E_{n,\bk,\sigma}$ is the corresponding quasiparticle excitation spectrum. The mean-field ground state is thus a free Fermi gas of Bogoliubov quasiparticles and we have
\bea
\langle \gamma^{\dagger}_{m\bk\sigma}\gamma_{n\bk'\sigma} \rangle_{\mathrm{MF}} = \delta_{mn}\delta_{\bk\bk'} n_F(E_{n\bk\sigma}) \non\\
\langle \gamma_{m\bk\sigma}\gamma_{n\bk'\sigma} \rangle_{\mathrm{MF}} = \langle \gamma^{\dagger}_{m\bk\sigma}\gamma^{\dagger}_{n\bk'\sigma} \rangle_{\mathrm{MF}} = 0 \,.
\eea
Here, $n_F(E_{n\bk\sigma})$ is the Fermi-Dirac distribution function defined as 
\beq
n_F(E_{n\bk\sigma}) = \frac{1}{1+e^{\beta E_{n\bk\sigma}}}\,
\eeq
with $\beta = 1/(k_B T)$. The relevant expectation values in the mean-field ground state can now be computed by writing the fermion operators in terms of the Bololiubov operators. For example,
\bea
\calF_{MF} &=& -k_B T \sum_{n,\bk,\sigma} log[1+e^{-\beta E_{n,\bk,\sigma}}].
\eea
The phase diagram obtained by direct minimization of the free energy in \eqn{eqn:fenergy1} is shown in the Fig. 2 of the main text. 

\begin{figure}
\includegraphics[width=0.999\linewidth]{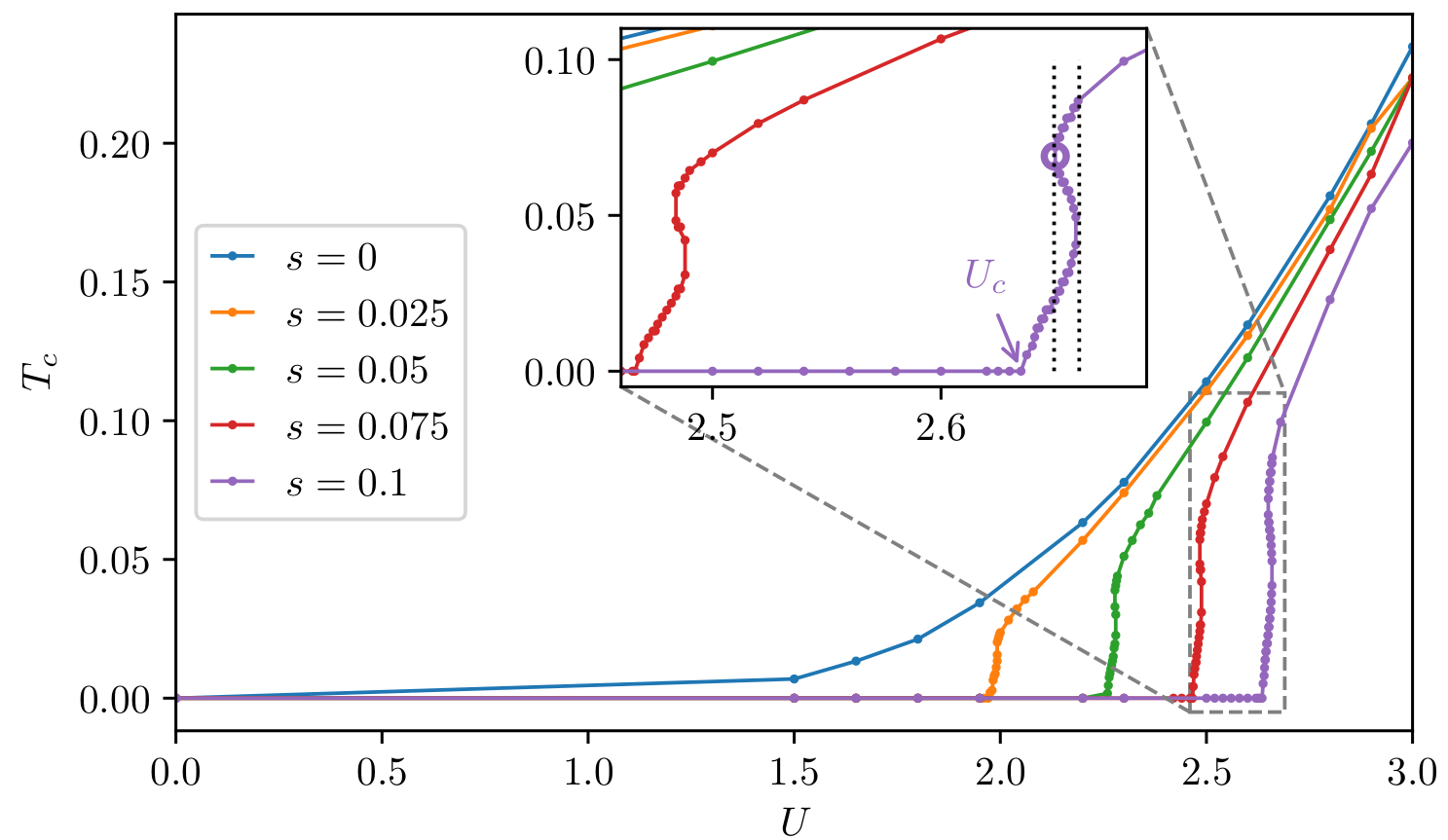}
\caption{\mylabel{fig:criticalU} Phase diagram obtained by direct minimization of the free energy using a hill descent algorithm. Each line marks the phase boundary between the normal state (higher T lower U side of line) and the superconducting state (lower T higher U side of line), with each color corresponding to different bare splitting $
s$. For finite splitting there exists a critical interaction $U_c$ (labeled for s = 0.1) below which superconductivity is fully suppressed. With finite splitting, just above $U_c$ there is a kink in the phase boundary caused by the bare splitting of the energy bands, leading to the possibility of re-entrant superconductivity. The circle marks
(shown for the s = 0.1 case only) the point where the phase boundary changes between first order (temperatures below this point) and second order (temperatures above this point).}
\end{figure}

Looking closely at the phase diagram as shown in \fig{fig:criticalU}, we note a kink (\fig{fig:criticalU} inset). This corresponds to re-entrant superconductivity. If the system has an interaction, U, between the two vertical dotted lines in the inset and is initially at T = 0 then it will be superconducting. As the temperature increases it will undergo a first order transition into the normal state. Further increase in T will then see another first order transition into a superconducting state, then followed by a further second order phase transition into the normal state. This is a form of order by disorder where the increase in temperature causes the stabilization of the more ordered superconducting state. The empty purple circle in the inset of \fig{fig:criticalU} marks the point where the phase boundary goes from a first order transition boundary (for temperatures below this point) to a second order transition boundary (for temperatures above this point). Although shown only for the s = 0.1 case, the same point is present in all phase boundaries for finite bare splitting $s$ and the same physics occurs however, the size of the re-entrant region decreases with bare splitting. We note that the re-entrant behavior is present only in a tiny portion of the phase diagram, over a small range of interaction ($2.65 \lesssim U \lesssim 2.66$ for the s = 0.1 case) which is related to the size of the bare splitting. Bearing in mind the band width in this model is 8t, the interaction range over which this can be observed is approximately 0.01t, or 800 times smaller than the band width. This only gets smaller as the splitting is decreased and, realistically, the bare splitting must not be too large otherwise superconductivity becomes improbable. This fascinating phenomenon will therefore be difficult to observe in practice.

\subsection{Details of the Dirac BdG calculations for LaNiGa$_2$}

The relativistic Kohn-Sham-Dirac BdG Hamiltonian is 
\beq
 \mathcal{H}_{DBdG} = 
  \left[ {\begin{array}{cc}
   H_D & \hat \Delta(\br)  \\
   \hat \Delta^*(\br)  & -H^*_D \\
  \end{array} } \right]
\eeq
where $\hat \Delta(\br)$ is the pairing potential matrix ($4\times 4$), and 
$H_D$ is the effective normal state Dirac Hamiltonian given by 
\beq
H_D = c \bp \hat{\pmb{\alpha}}_1 + (\hat{\pmb{\alpha}}_2 - \mathbb{1}_4)c^2/2 + (V_{eff}(\br) - E_F)\mathbb{1}_4 + \mathbf{B}_{eff}(\br)\hat{\pmb{\alpha}}_3.
\eeq
Here, $\hat{\pmb{\alpha}}_1 = \hat{\sigma}_x \otimes \hat{\pmb{\sigma}}$, $\hat{\pmb{\alpha}}_2 = \hat{\sigma}_z \otimes \mathbb{1}_2$ and $\hat{\pmb{\alpha}}_3 = \mathbb{1}_2 \otimes \hat{\pmb{\sigma}}$ with $\hat{\pmb{\sigma}}$ being the Pauli matrices and $\mathbb{1}_n$ being the identity matrix of order $n$.
$V_{eff}(\br)$ and $\mathbf{B}_{eff}(\br)$ are the effective electrostatic potential and the effective exchange-field, respectively:
\begin{subequations}
\begin{eqnarray}
  V_{eff}(\br)   &=&   \int \frac{\rho(\br')}{|\br - \br'|}  d\br' +
  \frac{\delta E^0_{xc}[\rho,\mathbf m]}{\delta \rho(\br)},\\
  \mathbf{B}_{eff}(\br)&=&  \frac{\delta E^0_{xc}[\rho,\mathbf m]}{\delta \mathbf m(\br)},
\end{eqnarray}
\end{subequations}%
where
$\rho (\br)$ is the charge density,
$\mathbf m (\br)$ is the magnetization density,
$E^0_{xc}[\rho,\mathbf m]$ is the usual (local spin density approximation)
exchange correlation energy for normal state electrons. No
symmetry of the resulting gap function is assumed as it
is the direct consequence of the assumed pairing model.

In the KKR method, the potential is treated in the so called muffin-tin approximation, i.e.\
the potential is written as a sum of single-domain potentials centered around each lattice site.
The nicest feature of KKR is that it allows direct access to the Green's function through
multiple scattering theory
\begin{equation}
	G(z)=(z-H_{DBdG})^{-1}=
	\begin{pmatrix}
		G^{ee}(z) & G^{eh}(z) \\
		G^{he}(z) & G^{hh}(z)
	\end{pmatrix}
\end{equation}
where $G^{ee}(z)$, $G^{hh}(z)$ and $G^{eh}(z)$ are the electron-electron, hole-hole and electron-hole Green's functions respectively. 

In the relativistic case we expand the solutions of the DBdG equations in the following form
\begin{equation}
 \Psi(z, \br)= \sum_{Q}
 \begin{pmatrix}
  g^e_{Q}(z,r) \chi_{Q} (\hat r) \\
  i f^e_{Q}(z,r)\chi_{\overline Q} (\hat r) \\
  g^h_{Q}(z,r) \chi^*_{Q} (\hat r) \\
  - i f^h_{Q}(z,r)\chi^*_{\overline Q} (\hat r)
 \end{pmatrix},%
\end{equation}%
\vskip 0.2in
\noindent
where $z$ is the complex energy, $Q=(\kappa,\mu)$ and $\overline{Q}=(-\kappa,\mu)$ are the composite indices for the spin-orbit ($\kappa$) and magnetic ($\mu$) quantum numbers;
$g^{e(h)}_{Q}(z,r)$ and $f^{e(h)}_{Q}(z,r)$ are the large and small components of the electron (hole) part of the solution, respectively.
The spin-angular function is an eigenfunction of the spin-orbit operator $ K = \boldsymbol \sigma L +\mathbb{I}$
\begin{equation}
 K \ket{\kappa \mu} = -\kappa \ket{\kappa \mu}.
\end{equation}

With integration over the angular parts and using the orthonormality
of the Clebsch-Gordan coefficients, and assuming that
the pairing potential does not couple between the large and the small component solutions,
the radial DBdG equations for arbitrary magnetic field (single-site problem) can be written as
\begin{widetext}
\begin{equation}
\begin{split}
&
\begin{pmatrix}
z + E_F & -i c \left( \frac{d}{d r} + \frac{1}{r} - \frac{\kappa}{r} \right) & 0 & 0 \\
-i c \left( \frac{d}{d r} + \frac{1}{r} + \frac{\kappa}{r} \right) & z + E_F + c^2 & 0 & 0 \\
0 & 0 & E_F -z & i c \left( \frac{d}{d r} + \frac{1}{r} - \frac{\kappa}{r} \right) \\
0 & 0 & i c \left( \frac{d}{d r} + \frac{1}{r} + \frac{\kappa}{r} \right) & -z + E_F + c^2
\end{pmatrix}
\begin{pmatrix}
g^e_{Q}(z,r) \\
i f^e_{Q}(z,r) \\
g^h_{Q}(z,r) \\
-i f^h_{Q}(z,r)
\end{pmatrix}
 = \\
&
\sum_{Q'}
\begin{pmatrix}
u^{++}_{Q Q'}(r) & 0 & \Delta_{Q Q'}(r)  & 0 \\
0 & u^{--}_{Q Q'}(r) & 0 & \Delta_{Q Q'}(r) \\
\Delta^*_{Q Q'}(r) & 0 & u^{++}_{Q Q'}(r)^* & 0 \\
0 & \Delta^*_{Q Q'}(r) & 0 & u^{--}_{Q Q'}(r)^*
\end{pmatrix}
\begin{pmatrix}
g^e_{Q'}(z,r) \\
i f^e_{Q'}(z,r) \\
g^h_{Q'}(z,r) \\
-i f^h_{Q'}(z,r)
\end{pmatrix},
\end{split}
\end{equation}
\end{widetext}
where
\begin{equation}
u^{++}_{Q Q'}(r)= V(r) + \sum_{i=x,y,z}\matrixelem{\chi_{Q}}{ \sigma_i B_i(r)}{\chi_{Q'}},
\end{equation}
\begin{equation}
u^{--}_{Q Q'}(r)= V(r) - \sum_{i=x,y,z} \matrixelem{\chi_{\overline Q}}{ \sigma_i B_i(r)}{\chi_{\overline{Q}'}},
\end{equation}
and $\Delta_{Q Q'}(r)$ is obtained by a Clebsch-Gordan transformation from an $(L,\sigma)$ basis ($L=(l,m)$)
of $\Delta_{L,\sigma;L',\sigma}(r)$, assuming interaction between different orbitals but between same spins. In this way we can introduce the orbitally antisymmetric equal spin-pairing.

Based on the solution of the single-site problem we can obtain the t-matrix, and
the Green's function can be constructed in exactly the same way as it is described in
Ref.~\onlinecite{Csire2018}.
Then, the self-consistent equations for the
charge densities and the pairing potential in an $L,s$
representation reads as
\begin{widetext}
 \begin{equation}
    \rho_{\sigma}(r_n) = 
      - \frac{1}{\pi} \int d  \varepsilon \int_{\mathrm{BZ}} d \bk  ~
	 ~\Im~\! Tr_L~\! \left[
	 f(\varepsilon )
	 G^{ee}_{L,\sigma;L',\sigma}(\varepsilon + i 0 , r_n, \bk) +	 
	 (1-f(\varepsilon ))
	 G^{hh}_{L,\sigma;L',\sigma}(\varepsilon + i 0 , r_n, \bk) \right],
  \end{equation}
 \begin{equation}
    \Delta_{L,\sigma;L',\sigma}(r_n) = 
      - U_{L,L'}
\frac{1}{2\pi} \int d  \varepsilon \int_{\mathrm{BZ}} d \bk  ~ (1-2f(\varepsilon ))
       ~\Im~\! F_{L,\sigma;L',\sigma}(\varepsilon + i 0 , r_n, \bk),
  \end{equation}
\end{widetext}
where 
$F_{L,\sigma;L',\sigma}(\varepsilon + i 0 , r_n, \bk)
=G^{eh}_{L,\sigma;L',\sigma}(\varepsilon + i 0 , r_n, \bk)
-G^{eh}_{L',\sigma;L,\sigma}(\varepsilon + i 0 , r_n, \bk)$
denotes the orbitally antisymmetric part of the off-diagonal
electron-hole part of the Green's function,
and $r_n$ refers to the radius within the muffin-tin potential of the lattice site $n$.

In practice, this method is implemented into a layered (2D) KKR code, and the
Ceperley-Alder  exchange-correlation functional was used in the calculations.
The radial DBdG equations are solved with a corrector-predictor method on a logarithmic scale.
For the energy integrations we used 36 energy points on the complex contour (semi-circle),
while the 2D Brillouin zone integrations were performed employing 500$\times$500 $k$-points.
The charge density is assumed to be the same as it is in the normal state in the first step, and
the starting value of the pairing potential is an $r$ independent, constant orbitally antisymmetric guess for
$\Delta_{L,\sigma;L',\sigma}$. 
The self-consistency was reached in both the charge densities $\rho_\sigma(r_n)$ and the pairing potentials 
$\Delta_{L,\sigma;L',\sigma}(r_n)$.

The crystal structure of LaNiGa$_2$ is orthorombic with space group 65, {\it{Cmmm}}
and two formula units (8 atoms) per unit cell.
We used the experimentally reported lattice parameters~\cite{Romaka1983}
$a=4.29$~\AA,
$b=17.83$~\AA, 
$c=4.273$~\AA, 
with the following Wyckoff positions obtained by total energy minimization
La[4j](0, 0.359, 0.5), Ni[4i] (0, 0.072, 0), Ga[4i] (0, 0.21, 0), Ga[2d] (0,0,0.5), Ga[2b](0.5,0,0).

\begin{figure}[t!]
\includegraphics[width=0.98\linewidth]{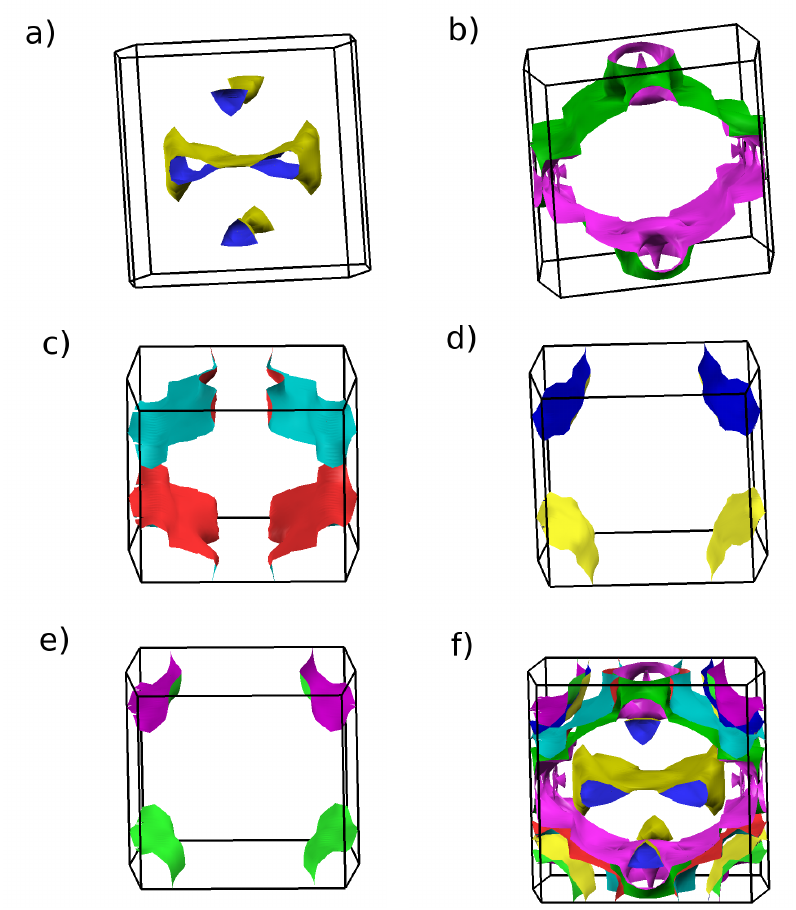}
\caption{\label{fig:fermi} Fermi surfaces of LaNiGa$_2$ with spin-orbit coupling. The panels (a)--(e) show the five Fermi surfaces corresponding to the five bands crossing the Fermi level. Panel (f) shows their topology after merging them all together.}
\end{figure}

\begin{figure}[hbt!]
\includegraphics[width=0.98\linewidth]{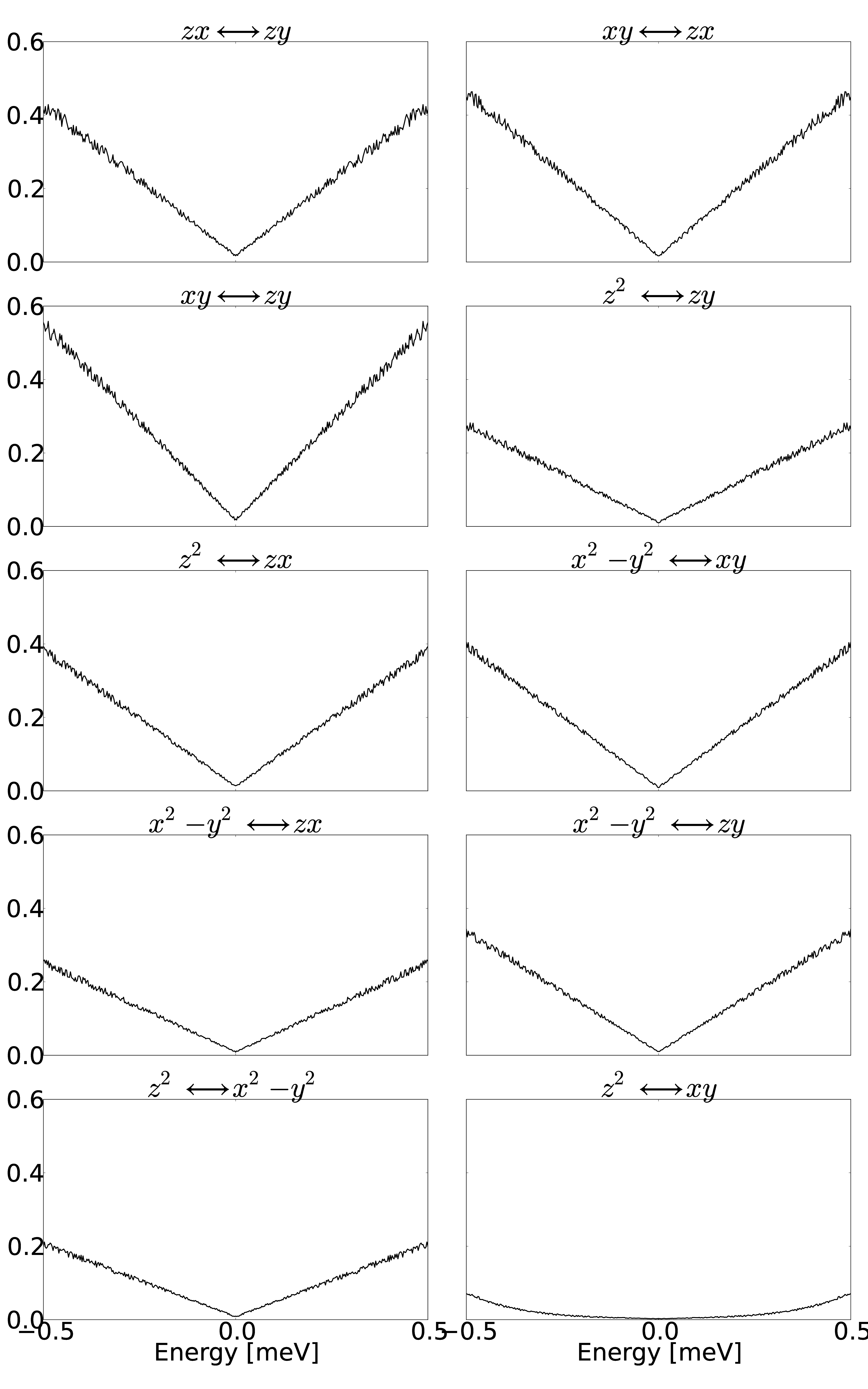}
\caption{\label{fig:qdos} Density of states (arbitrary units) in the superconducting state for different inter-orbital equal-spin pairing models on the $d$ orbitals of the Ni atom.}
\end{figure}

\subsection{Fermi surfaces and quasiparticle DOS of LaNiGa$_2$}
We have calculated the Fermi surfaces of LaNiGa2, as shown in Fig.~\ref{fig:fermi}, which were calculated using the fully relativistic bands. No significant differences were found compared to the Fermi surfaces obtained neglecting spin-orbit coupling (SOC) by using scalar relativistic bands. The fact that no significant change is found resulting from SOC is reasonable due to the centrosymmetric structure of the material.  These results agree well with the Fermi surface presented in Ref.~\cite{Singh2012}.

At the Fermi level, the density of states (DOS) contributions coming from all 5 Ni $d$-orbitals are relevant, leading to 10 possibilities for the inter-orbital equal-spin pairing model on the Ni atom. Therefore, by employing the DBdG-KKR method 
described in the previous section and assuming the superconducting gap initially to be 1 meV, we calculated the quasiparticle DOS to find which pairing model produces a fully gapped quasiparticle spectrum and which shows a nodal gap structure.
As shown in Fig.~\ref{fig:qdos}, most of the cases lead to a V-shaped DOS around the Fermi level implying line nodes in the quasiparticle spectrum. In fact, we have five Fermi-surface sheets and the V-shape behavior of the DOS indicates that at least one of them has a nodal gap structure. These slopes also determine the low-temperature specific heat as a function of temperature. However, equal-spin pairing on the d$_{z^2}$--d$_{xy}$ orbitals yields anisotropic fully gapped quasiparticle spectrum (on all of the Fermi-surface sheets). Since experiments~\cite{Chen2013,Chen2013A,Weng2016} suggest fully gapped behavior in this material, we have chosen this scenario for our further investigations and results are presented in the main text. 

\bibliography{ESP_Mag_v03}

\end{document}